\documentclass[aps,prd,twocolumn,superscriptaddress,longbibliography,nofootinbib]{revtex4-2}
\usepackage[utf8]{inputenc}
\usepackage{color}
\usepackage{graphicx}
\usepackage{subfigure}
\usepackage{xcolor}
\usepackage[normalem]{ulem}
\usepackage{hyperref}
\usepackage{orcidlink}
\usepackage{float}
\usepackage{multirow}
\usepackage{hhline}
\usepackage{amsmath,amssymb,amsfonts}

\def\prl{Phys. Rev. Lett.}
\def\prd{Phys. Rev. D}
\def\cqg{Class. Quantum Grav.}

\def\pau_p{Prog. Theor. Phys.}

\def\prague{{\tt prague}}
\def\sphGR{{\tt sphGR}}
\def\bamps{{\tt bamps}}

\def\p{\partial}

\definecolor{cyan_dp}{HTML}{007272}
\definecolor{rose}{HTML}{a84847}
\definecolor{green}{HTML}{008060}
\definecolor{orange}{HTML}{FF6600}
\definecolor{red}{HTML}{FF2511}
\definecolor{blue}{HTML}{055C9D}
\definecolor{orchid}{HTML}{9c3386}
\definecolor{ref_blue}{HTML}{00478F}

\hypersetup{
    colorlinks,
    citecolor=blue,
    filecolor=black,
    linkcolor=green,
    urlcolor=teal
}

\newcommand{\ds}{\displaystyle}

\begin{document}

\title{Comparing twist-free axisymmetric gravitational waves near the
  black hole threshold}

\author{Ananya Adhikari\orcidlink{0000-0002-3890-3577}}
\affiliation{
  Centro de Astrof\'{\i}sica e Gravita\c c\~ao -- CENTRA, Departamento
  de F\'{\i}sica, Instituto Superior T\'ecnico -- IST, Universidade de
  Lisboa -- UL, Av.\ Rovisco Pais 1, 1049-001 Lisboa, Portugal}
  
\author{Tom\'a\v{s} Ledvinka\orcidlink{0000-0002-6341-2227}}
\affiliation{Institute of Theoretical Physics, Faculty of Mathematics
  and Physics, Charles University, CZ-180 00 Prague, Czech Republic }

\author{Daniela Cors\orcidlink{0000-0002-0520-2600}}
\affiliation{Department of Applied Mathematics and Theoretical
  Physics, Centre for Mathematical Sciences, University of Cambridge,
  Wilberforce Road, Cambridge CB3 0WA, United Kingdom}

\author{Thomas~W.~Baumgarte\orcidlink{0000-0002-6316-602X}}
\affiliation{Department of Physics and Astronomy, Bowdoin College,
  Brunswick, ME 04011, USA}

\author{Anton Khirnov\orcidlink{0000-0002-9569-381X}}
\affiliation{Institute of Theoretical Physics, Faculty of Mathematics
  and Physics, Charles University, CZ-180 00 Prague, Czech Republic }

\author{Bernd Brügmann\orcidlink{0000-0003-4623-0525}}
\affiliation{Friedrich-Schiller-Universität, Jena, 07743 Jena,
  Germany}
  
\author{David~Hilditch\orcidlink{0000-0001-9960-5293}} \affiliation{
  Centro de Astrof\'{\i}sica e Gravita\c c\~ao -- CENTRA, Departamento
  de F\'{\i}sica, Instituto Superior T\'ecnico -- IST, Universidade de
  Lisboa -- UL, Av.\ Rovisco Pais 1, 1049-001 Lisboa, Portugal}

\begin{abstract}
  The threshold of black hole formation in axisymmetric vacuum gravity is proving to be more complicated than had been anticipated but, following recent advances, a consensus between independent codes and methods is emerging. Building on earlier work we provide further details of a comparison between three independent numerical codes (the \bamps, \prague, and~\sphGR~codes), paying special attention to the relative strengths and weaknesses of each and examining various features of near-threshold collapse of vacuum gravitational waves for the first time. In particular, we observe quasi-universal strong-field features appearing in curvature scalars. Focusing on geometric features on the symmetry axis, we construct reference coordinates to aid the comparison of strong-field data. We evolve, for the first time, time-asymmetric wave initial data within the \bamps~code. To the extent possible with current methods we compare apparent horizons and attempt to determine what causes difficulties in the classification of these strong-field and highly dynamical spacetimes. In all cases the results from the three codes agree very well.
\end{abstract}

\maketitle

\section{Introduction}
\label{sec:Introduction}

It has long been known~\cite{Cho93,GunM07,GunHilMar25} that solutions of general relativity (GR) close to the threshold of black hole formation exhibit interesting properties, commonly referred to as critical phenomena in gravitational collapse. Strong numerical evidence suggests that, at least in the context of spherical symmetry, these properties, such as power-law scaling of the maximum value of curvature scalars, are associated with the emergence of a universal
self-similar solution precisely at the threshold.

Yet, the situation is less clear-cut as the degree of symmetry is reduced. In spacetimes with three spatial dimensions the easiest way to relax spherical symmetry is to consider axisymmetry.  Of particular interest are axisymmetric gravitational waves, since they probe the characteristics of GR alone, independently of those of other matter models. Employing the standard strategy of bisection within one-parameter families, such waves have been brought close to the black hole threshold in the pioneering work of Abrahams and Evans in the early 1990s~\cite{AbeE93,AbeE94}. Subsequent work, however, failed to recover the phenomenology -- the existence of an assumed universal self-similar critical solution with a unique critical exponent -- suggested by this early breakthrough. Worse, some later numerical studies  reported different critical exponents and hence were (occasionally) seemingly in outright contradiction with each other.

Some of these puzzles have been resolved in a series of recent papers~\cite{LedK21,SuaRCBH22,BauGH23,BauBC23}. In particular, it now appears that there is no unique critical solution for vacuum collapse, at least at the current level of fine-tuning, so that different families of initial data may result in different properties close to the threshold of black hole formation. In recent work~\cite{BauBC23} we have demonstrated that the independent \bamps, \prague, and \sphGR~codes give quantitatively compatible results, providing confidence that the above conclusions are not based on numerical artifacts. Considering two families of Brill wave (time-symmetric) initial data, we found approximate power-law scaling in the maxima of curvature scalars but with family-dependent exponents. We found that in one of the families the peak of the curvature, typically found on the symmetry axis away from the origin, was accumulating closer to the origin as we tuned closer to the threshold of collapse. Introducing single-null coordinates adapted to self-similarity centered at the origin within this particular family, we found evidence for approximate self-similarity, with near perfect agreement between the numerical data. This is a particularly hard comparison because it occurs over an extended region of a strong-field and highly dynamical spacetime.

In this paper, we return to the problem with the aim of giving an overview of progress on the topic, more details of our earlier comparison, and to extend by considering aspects of the spacetimes that we had not analyzed before, increasing furthermore the number of spacetimes treated in common with the different codes. Specifically, we now evolve off-center Brill waves (see \cite{SuaRCBH22}) with the \prague~code, as well as nonlinear time-asymmetric waves (using the initial data of~\cite{LedK21}) within \bamps. To the extent that we are able to tune to the threshold with each method we find remarkable agreement across the entire spectrum of families considered. This includes the parametric dependence of the spacetimes as we move through the solution space (for instance in curvature scaling), the representation of specific spacetimes (for instance once they are re-expressed in canonical single or double-null coordinates) and, far from the black hole threshold, apparent horizon masses. This gives further confidence to the physical interpretation arrived at in our previous work~\cite{BauBC23}.

The crux of work in this area is the degree of tuning to the black hole threshold that can be attained. Axisymmetric vacuum spacetimes have proven challenging in this regard. To give a sense of this, consider the fact that many spherical studies regularly arrive at a tuning within a one parameter family to double-precision accuracy, whereas here, even with the present state-of-the-art, we achieve at best seven decimal places and only for certain families. As discussed in detail below, the comparison here highlights the relative strengths and weaknesses of different approaches and thus points the way for improvements in future work.

The paper is structured as follows. First, in section~\ref{sec:Methods} we give a brief overview of the geometric setup, our classes of initial data, numerical tools and approach to the threshold. In section~\ref{sec:Results} we then present and discuss our results. Section~\ref{sec:Conclusions} contains our conclusions. We work in geometric units with~$G=c=1$ throughout.

\section{Methodology overview}
\label{sec:Methods}

\begin{table*}[t]
  \centering
  \begin{ruledtabular}
  \begin{tabular}{l|l|l|l|l|l}
    \hline
Authors & Year & Data type  & Setup & Ref.  & Comments \\
\hline
Eppley & 1978 & Brill & maximal slicing/quasiisotropic & \cite{Epp79} & Small amplitude waves only \\
Miyama & 1980 & Brill/Time-asymm. & Geodesic slicing & \cite{Miy80} & \\
Abrahams \& Evans & 1992 & Teukolsky & maximal slicing/quasiisotropic & \cite{AbeE93,AbeE94} & Reported critical behavior \\
Alcubierre {\it et.al.} & 2000 & Brill  & maximal slicing/zero shift & \cite{AlcABLSST00} \\
Garfinkle \& Duncan & 2001 &Brill & maximal slicing/quasiisotropic & \cite{GarD01} \\
Rinne & 2008 & Brill & maximal slicing/quasiisotropic & \cite{Rin08a} \\
Sorkin & 2011 & Brill & GHG & \cite{Sor10a} & \\
Hilditch \it{et. al.} & 2013 & Brill/Teukolsky & Moving-puncture/1+$\log$, zero shift & \cite{Hiletal13} & Coordinate singularities \\
Hilditch \it{et. al.} & 2017 & Brill & GHG, damped gauge sources & \cite{HilWB17} & Curvature scaling, bifurcation \\
Ledvinka \& Khirnov & 2021 & Brill/Time-asym. & Quasi-maximal, zero shift & \cite{LedK21,Khi21} & Non-universal scaling, echoes \\
Su\'arez Fern\'andez  \it{et. al.} & 2022 & Off-center Brill & GHG, damped gauge sources & \cite{SuaRCBH22} &
Non-universal scaling, echoes \\
Baumgarte \it{et. al.} & 2023 & Teukolsky, $l=2,4$ & Shock-avoiding, zero shift & \cite{BauGH23} &
Non-universal scaling, echoes \\
Baumgarte \it{et. al.} & 2023 & Brill & Various: comparison paper & \cite{BauBC23} & Agreement in single-null\\
Baumgarte \it{et. al.} & 2026 & Nakamura  &  Shock-avoiding, zero shift & \cite{BauGunHil26} & Non-universal scaling, echoes\\
\hline
\end{tabular}
  \end{ruledtabular}
  \caption{Overview of results on numerical simulations of axisymmetric twist-free gravitational waves. Updated from that in~\cite{Hiletal13}. We observe that while early results vary in their scientific goals, the degree of tuning achieved, and especially in their conclusions, recent works are consistent in their findings. In the direct precursor to this work~\cite{BauBC23} a quantitative comparison was made between Brill wave data as treated in the \bamps, \prague, and \sphGR~codes.}
\label{Table:PreviousResults}
\end{table*}

\subsection{Geometric setup and initial data}
\label{sec:Geometry_ID}

We work in vacuum in twist-free axisymmetric GR, with metric~$g_{ab}$
and covariant derivative~$\nabla_a$. This means that we solve
numerically the initial value problem for the vacuum Einstein
equations
\begin{align}
  G_{ab}&=8\pi T_{ab} =0\,,
\end{align}
thus constructing the metric~$g_{ab}$. We use the standard numerical
relativity terminology for the lapse~$\alpha$, shift~$\beta^i$ and
spatial metric~$\gamma_{ab}$. Twist-free axisymmetry means that there
exists a Killing vector~$\xi^a=\partial_\varphi{}^a$ with closed
orbits such that~$\epsilon_{abcd}\xi^b\nabla^c\xi^d=0$. Within this
class, we consider two types of initial data, with different
approaches to the solution of the vacuum Hamiltonian and momentum
constraints. In Table~\ref{Table:PreviousResults} we give an overview of earlier numerical evolutions of such initial data.

\paragraph*{Brill waves.} Brill wave data~\cite{Bri59} are constructed by choosing a moment of time symmetry~$K_{ij} = 0$ as well as a conformally related spatial metric~$\tilde \gamma_{ij} = \psi^{-4} \gamma_{ij}$ of the form
\begin{align}
  \tilde{\gamma}_{ij}\text{d}x^i\text{d}x^j
  =e^q(\text{d}\rho^2+\text{d}z^2)+\rho^2\text{d}\varphi^2\,,
\end{align}
where we adopted cylindrical polar coordinates. Choosing a seed-function~$q$ and inserting these choices into the Hamiltonian and momentum constraints gives a linear elliptic equation for the conformal factor~$\psi$ (see~\cite{HilWB16,KhiL18} for details of the numerical approach). Specifically, we choose
\begin{align}
  q&=A\rho^2e^{-[(\rho-\rho_0)^2/\sigma_\rho^2+(z-z_0)^2/\sigma_z^2]}\,.
  \label{eq:Brill_seed}
\end{align}
Fixing~$\rho_0$, we then tune the amplitude parameter~$A$ to the
threshold of collapse as discussed below. In~\cite{BauBC23}, we considered two families ($A>0$ and~$A<0$) of centered ($\rho_0=0$) Brill waves. Here we extend that work by considering also off-center Brill wave data with~$\rho_0=5$, as evolved in~\cite{SuaRCBH22} with the \bamps~code, but now treated also in the \prague~and \sphGR~codes.

\paragraph*{Time-asymmetric waves.} The main focus of this study is to
compare the evolution of nonlinear initial data based on Teukolsky waves \cite{Teu82}. The basic idea of these initial data is to take solutions to the
linearization of vacuum GR around flat-space~\cite{Teu82,Rin08b} and `dress' them in such a way as to solve the nonlinear constraints. The second of these steps can be done in rather different ways, see for instance~\cite{Nak84,ShiN95,Hiletal13,BauGH23}. Our approach is that introduced in~\cite{LedK21}, which is slightly different from that of Abrahams and Evans~\cite{AbeE93} (see~\cite{Ros25} for a variation that is likely more faithful to their original formulation). Ultimately we choose a conformally flat metric and maximal slicing~$K=0$ initially, and set the extrinsic curvature
component~$K_\theta{}^r$ to be
\begin{align}
K_\theta{}^r&=
A\sigma^{-4}r^2(\sigma-r)e^{-r^2/\sigma^2}\sin2\theta
\label{eq:Teukolsky_seed}
\end{align}
in spherical polar coordinates. Placing these into the Hamiltonian and
momentum constraints yields a coupled set of partial differential equations (PDEs) for the conformal factor~$\psi$ and the two extrinsic curvature components~$K_r{}^r$ and $K_\varphi{}^\varphi$. We solve these equations as in the case of Brill waves by imposing asymptotic flatness at spatial infinity. It turns out that there are two branches of solutions to this formulation of the constraints.  We denote the solutions in the upper (higher mass) branch with an overline, for instance~$A = \overline{1.3}$. For further details, especially as concerns non-uniqueness of solutions, see~\cite{LedK21,Khi21}.

\subsection{Evolution codes}\label{sec:Code}

We work with three independent numerical relativity evolution codes.

\paragraph*{\bamps\!.} The \bamps~code~\cite{Bru11,HilWB16,RenCH23, CorRenRue23} employs an adaptive multidomain pseudospectral method to evolve symmetric hyperbolic systems of PDEs under the method of lines in 3+1 dimensions. For the purposes of computational efficiency in the context of axisymmetry, we employ the cartoon method as formulated in~\cite{Pre04}. This allows us to evolve on a 2+1 dimensional domain, with transverse derivatives evaluated using the Killing vector. We work with a first order generalized harmonic formulation of GR~\cite{LinSKOR06}. Data are communicated between neighboring spectral elements via a penalty method. The outer boundary is dealt with using constraint-preserving outer boundary conditions. Mesh refinement dynamically adjusts the approximation to spatial derivatives in two distinct ways. First, smaller spectral elements can be introduced or removed as needed (h-refinement) and, second, the polynomial order of the approximation to spatial derivatives is dynamically adjusted (p-refinement)
~\cite{RenCH23}. The decisions for this are made by estimates of truncation error or the smoothness of the data on the domain. In practice, in the runs presented in this work we employ only~$h$-refinement. The code is parallelized with MPI and displays convincing strong scaling on at least several thousand cores.

\paragraph*{\prague\!.} The \prague~code~\cite{Khi21} is built within
the Cactus computational environment~\cite{cactusweb1} and employs several
components from the Einstein Toolkit~\cite{ET12,ZilLoe13,einsteintoolkitweb}. It solves hyperbolic
formulations of GR with the method of lines, using finite differencing
for spatial derivatives. Fixed box-in-box mesh refinement (FMR) allows for
the accurate resolution of a desired region (around the origin in the
work we discuss here), and the cartoon method is once more used to
reduce the numerical domain to 2+1. Outer boundaries are causally disconnected from the region of interest for the duration of the simulations. For the time evolution itself, the code employs the BSSN
formulation~\cite{NakOK87,ShiN95,BauS98}
together with the `quasi-maximal' slicing condition introduced in~\cite{KhiL18} and a
vanishing shift.

\paragraph*{\sphGR\!.} The \sphGR~code uses spherical polar
coordinates~\cite{BauMCM13,BauMonMue15} and eighth-order spatial
finite differencing in conjunction with a fourth-order method of lines for time evolution. Spherical polar coordinates are well-adapted to situations
in which there is a single center of collapse at the origin, and also
have the advantage that axisymmetric problems can trivially be treated
with a single point in the~$\phi$ direction.  Singular terms
that appear in these coordinates are managed by using a cell-centered
mesh and decomposing the evolved fields against unit vectors
normalized within a reference-metric formulation. The code also employs an asymptotically logarithmic radial coordinate and, furthermore, contracts the grid successively to achieve greater resolution in the region of interest. \sphGR~evolves the BSSN formulation using a shock-avoiding~\cite{Alc97,Alc02,BauH22,BeiChrWan25} variant of the Bona-Masso slicing condition~\cite{BonMas94} together with a vanishing shift.

\subsection{Bisection and classification}\label{sec:Bisection}

Our general methodology is to fix a one-parameter (often the amplitude
of the initial data) family of initial data and bisect the parameter
within that family to the threshold of black hole formation. Ideally
this means running simulations long enough until one of two things
happens. Either the data disperse and so can be classified as
subcritical, or a black hole is formed and so they are supercritical.

Subcritical data are characterized by late-time decay of curvature scalars, such as the Kretschmann scalar
\begin{align} \label{eq:Kretschmann}
I=R_{abcd}R^{abcd}.
\end{align}
Specifically, we can identify a spacetime as subcritical if, after a sufficiently long evolution, the Kretschmann scalar has dropped off to a value that is orders of magnitude smaller than its absolute maximum value attained during the course of the evolution, and continues to decrease.

Classifying supercritical data is more involved. In our present approaches, supercritical data will eventually cause the codes to crash. But crashes obviously do not necessarily imply black hole formation. Numerical evolution could instead fail because of large error, which, close to the threshold, often manifests itself as high-frequency oscillations. We therefore confidently classify spacetimes as supercritical when we can search and find apparent horizons, as discussed momentarily, within the data. In practice, this step is subtle, and consequently failure to find a horizon is often the termination condition for the bisection. In cases in which we fail to find an apparent horizon, we can revisit the numerical setup to find a better solution for the spacetime by varying the combination of constraint damping, gauge, resolution and so forth and repeat. In cases that we can then successfully classify, we have to consider the fact that, numerically, the exact critical amplitude will depend on these changes, and thus we may have to rerun very tuned simulations (in order to rewind the bisection). This process is then followed iteratively to narrow down the difference between the weakest supercritical and the strongest subcritical initial data amplitudes.

\subsection{Apparent horizon finders}
\label{sec:AHFinders}

A smooth closed 2-surface~$S$ embedded in a spatial slice~$\Sigma_t$,
is called a marginally outer trapped surface (MOTS) when the
expansion~$\Theta$ of the outgoing null geodesics vanishes,
\begin{equation}
  \Theta = \nabla_a l^a = 0, \label{eq:expansion}
\end{equation}
with~$l^a=s^a+n^a$ being the null-like tangent to~$S$ up to an
arbitrary normalization factor, composed of the space-like~$s^a$
outward pointing normal to~$S$ tangent to~$\Sigma_t$, and the
time-like normal to~$\Sigma_t$, $n^a$. By definition, the outermost
MOTS is called the apparent horizon (AH). The AH is a quasilocal
(local in time) proxy for the presence of a black hole but is,
unfortunately, foliation dependent. (See~\cite{Alc08,BauSha10} for
textbook discussion in the context of numerical relativity).  Finding
the AH then reduces to prescribing methods for finding the outermost
2-surface on which Eq.~\eqref{eq:expansion} holds. We use two such
prescriptions, each with its own nuances, that are briefly described
below. For a broad overview of horizon searching methods, see~\cite{Tho06, Alc08, BauSha10}

\paragraph*{Shooting method:} In this approach, we take the
associated PDE~\eqref{eq:expansion} and reduce explicitly to
axisymmetry, which produces an ODE. We then `shoot' (hence the
name) curves normal to the symmetry axis by integrating this
equation. The resulting set of curves is then inspected to identify
any that turns back towards the symmetry axis and hits it at
right angles. By symmetry, this ensures that this forms a smooth
closed 2-surface. Our current finder uses a uniform grid to construct
the these curves. This means, it can directly use the fields
from the native uniform grid of single refinement level of \prague,
but needs to interpolate the adaptively refined grid of \bamps~or the
FMR hierarchy of \prague~to a (usually downsampled) uniform grid. While this means that the process is computationally cheaper, it potentially can lose information about the features in the fields. Solution curves to the ODE are often parametrized in such a way that only `star-shaped' surfaces can be represented but, as discussed
in~\cite{PooBirKri18} and~\cite{Khi21} with our implementation, this
can be avoided by use of a reference surface. In comparison with the
flow-method discussed below, the shooting method thus has the
advantage that it does not place assumptions on shape of the AH, and
hence is more robust at locating the deformed AHs that form close to
the threshold of collapse. Interestingly, we shall see that this method
is also capable of finding a marginally inner trapped surface
(MITS). This also means that the present version of this finder does
need some hand-holding as it may otherwise get trapped inside the event horizon in a region where the solution diverges, when fully automated.

\paragraph*{Flow method:}
In this alternative prescription, we
start with an initial guess for the 2-surface as a coordinate 2-sphere
on the slice of interest. The expansion is then computed on this
surface and the surface is made to `flow' (basis of the name)
towards the region where the expansion vanishes by using the
information from the fields. The advantage of this method is that it
requires very little hand-holding. Our implementation works with the
information of the native grid of \bamps, the evolution code employed
with the most complicated grid hierarchy. However, this also means
that it is computationally costly, making it slow. The flow method
additionally does assume the AH is star-shaped, and hence struggles,
or fails, to find, more deformed AHs. Empirically we find that the
method is quite sensitive to the initial guess.

\section{Results}
\label{sec:Results}

\begin{figure*}[t!]
\includegraphics[width=\linewidth]
{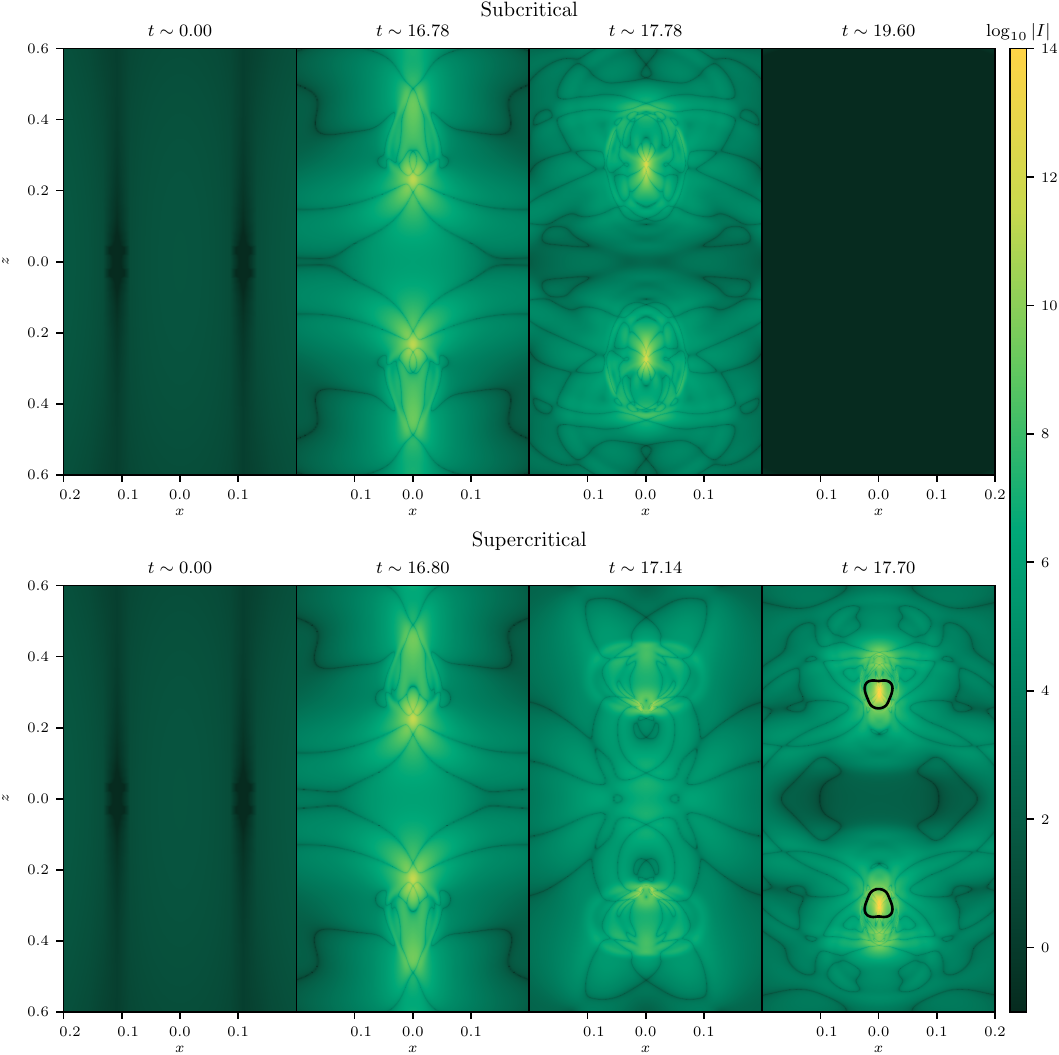}
\caption{\label{fig:Kretch2D} Top: The Kretschmann scalar (\ref{eq:Kretschmann}) from \bamps~in
  the evolution of dispersing time-asymmetric initial data with
  amplitude~$A=\overline{1.30080779}$ at coordinate times~$t \sim
  0.0,16.78,17.78,19.60$. After a number of strong oscillations on the
  symmetry axis the data rapidly disperse. Bottom: The Kretschmann scalar
  from \bamps~in the evolution of non-dispersing time-asymmetric
  wave initial data with amplitude~$A=\overline{1.300807615}$, at
  coordinate times $t \sim 0.0,16.80,17.14,17.70$. The strong
  oscillations keep growing in magnitude until apparent horizons form, which are shown in the bottom right panel for~$t \sim 17.70$ as the solid black lines.}
\end{figure*}

In this section we present a comparison of numerical solutions of both
Brill and time-asymmetric waves obtained with our three evolution
codes. We begin in section~\ref{sec:Basic} with an overview of the
basic dynamics of strongly gravitating vacuum spacetimes. In
section~\ref{sec:Parametric} we examine the scaling properties of
curvature on the subcritical side and apparent horizon mass for
supercritical spacetimes. In section~\ref{sec:Specific} we look in
more detail at specific spacetimes close to the threshold of collapse.

\subsection{Basic dynamics}
\label{sec:Basic}

\begin{figure*}[t!]
\includegraphics[width=\linewidth]{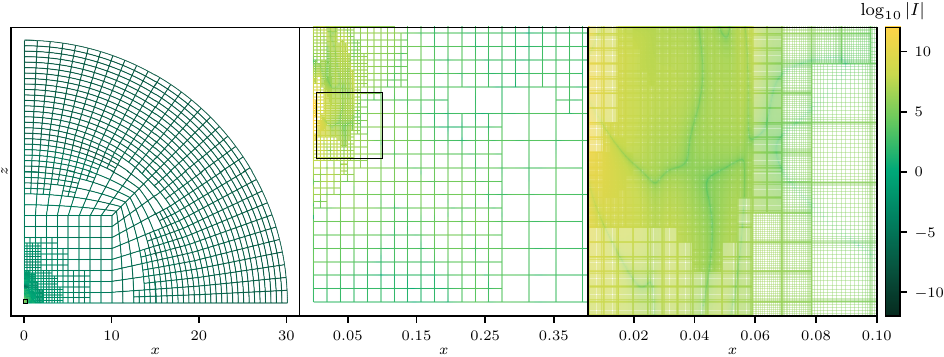}
\caption{\label{fig:KretchGrid} The Kretschmann Scalar on \bamps~grid
  for~$A=\overline{1.3008077675}$, the most tuned sub critical
  time-asymmetric data, at coordinate time~$t=17.78$. In the left
  most plot, we see the whole computational domain. In the center we
  show the tiny region marked by the square in the left most plots
  lower left corner, where the strongest dynamics occur. The right
  hand panel in turn shows the zoomed in view of the region in the
  black box in the central panel. The left and middle figures show
  only the domain boundaries while the right most figures shows the
  whole structure including the Gauss-Lobatto gridpoints within the
  domains. This last panel also shows regions with the finest
  resolution in the entire domain. The scale in each panel
  is shown along the $x$-direction and the extent is the same
  in both directions for each individual plot.}
\end{figure*}

Consistent with the nonlinear stability of the Minkowski spacetime,
numerical evolutions of weak data disperse, and this can be observed
robustly with any of our three methods. Near-critical data display violent
oscillations of curvature scalars, which are particularly visible on
the symmetry axis. When treating sufficiently large data we observe
that curvature scalars eventually diverge after an AH forms. As a
representative example, in Fig.~\ref{fig:Kretch2D} we plot snapshots
of the Kretschmann scalar~\eqref{eq:Kretschmann}) from evolutions of dispersing and collapsing time-asymmetric data as computed with~\bamps. Each code employs a
slightly different approach to maintaining enough resolution in the
strong-field region to accurately resolve these oscillations. To give a sense
of the grid hierarchy obtained under the~\bamps~adaptive mesh scheme,
in Fig.~\ref{fig:KretchGrid} we present nested snapshots of the grid
at a particular instance of time extracted from the subcritical
data. In our \bamps~runs, we allow for up to~$15$ levels of~$h$-refinement, though in practice arrive at most to~$12$ in the case shown. \prague~instead employs a nested fixed mesh approach, typically
with around~$11$ levels of~$2:1$ refined grids, while \sphGR~periodically
regrids its asymptotically logarithmic radial grid onto a smaller domain
to increase resolution around the origin.

\begin{figure}[t!]
\includegraphics[width=\linewidth]{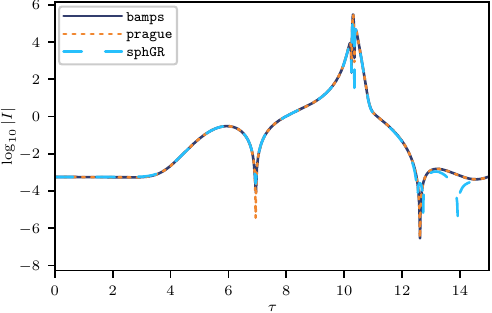}
\caption{\label{fig:BrillOffCent} The invariant~$I$ at the origin for
  Brill wave data with~$A = -0.047$ and~$z_0=0, \rho_0=5$, as computed with the \bamps, \prague, and \sphGR~codes.  At late times, numerical error at the origin becomes increasingly noticeable in the \sphGR~code, which employs spherical polar coordinates. Nevertheless the agreement between all three codes is good here.}
\end{figure}

As a first concrete point of comparison we examine the Kretschmann scalar~$I$ (\ref{eq:Kretschmann}). In Fig.~\ref{fig:BrillOffCent} we plot~$I$ at the origin as a function of proper time in the evolution of~$A=-0.04700$ off-center~$z_0=0,\rho_0=5$, $\sigma_z=\sigma_\rho=1$ Brill wave initial data as treated in~\cite{SuaRCBH22}, finding excellent agreement.

\subsection{Parametric dependence}
\label{sec:Parametric}

\begin{figure}[t!]
\includegraphics[width=\linewidth]{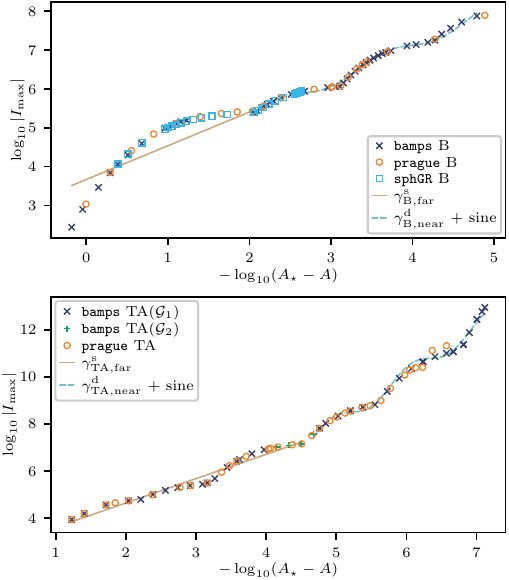}
\caption{\label{fig:Scaling} Scaling of the Kretschmann scalar in \bamps, \prague, and \sphGR~evolutions of the Brill wave family built from~\eqref{eq:Brill_seed} with $A<0$ (B, top panel), and the (upper branch) time-asymmetric initial data built from the
  ansatz~\eqref{eq:Teukolsky_seed} (TA, bottom panel, \bamps~and \prague~only), both on the subcritical side. The agreement among the codes is excellent throughout almost the
  entire range of tuning to the black hole threshold. As discussed in detail in the main text, in the \bamps~runs we needed to evolve with an alternative gauge for a small interval of initial amplitudes to avoid coordinate singularities. These are indicated with green `+' markers rather than blue `x' markers. While it is true that the slope of the curve
  appears to change in the last few points, obtained only with \bamps,
  these spacetimes are closest to the threshold and thus more strongly
  affected by the uncertainty in the critical amplitude (we estimate
  the critical amplitude as the mid-point of our best tuned sub- and
  supercritical data). The fits were performed on the \bamps~data and the sinusoidal piece in both the B and TA cases are added with the `near' region double fit $\Delta$ (see Table~\ref{Table:Fits}).}
\end{figure}

Much of our understanding of critical collapse comes not from studying specific highly tuned spacetimes, but rather the dependence of spacetime against the parameter we vary towards the threshold of collapse.

In the spherically symmetric setting, evidence suggests that, for specific fixed matter models, there exists a unique self-similar solution, called the critical solution, that lies at the threshold of collapse. Strong heuristic arguments suggest that such critical solutions are codimension-$1$ attractors with a single unstable mode with Lyapunov exponent~$\lambda_0$. General arguments~\cite{GarD98,GunHilMar25} then suggest that maxima of curvature invariants~$\mathcal{C}$ with units of length should scale according to a universal power-law~$\mathcal{C}\simeq (A-A_\star)^{-\gamma}$, where the critical exponent~$\gamma=1/\lambda_0$, as we approach the threshold from the subcritical side. If the critical solution is discretely-self-similar (DSS), this power-law will be superposed with a universal periodic function~\cite{Gun97,HodP97} (see the discussion below).

In axisymmetry the degree of tuning to the threshold is generally
rather less exquisite than in the spherical setting, so that all conclusions drawn from these simulations are subject to the disclaimer `at the current level of fine-tuning'.  Nevertheless, a number of different numerical experiments have revealed that, in the absence of spherical symmetry, the threshold of black-hole formation is significantly more complicated than in spherical symmetry.  For example, various studies of axisymmetric, minimally coupled massless scalar fields report curvature maxima on the symmetry axis rather than at the center, and suggest that, while power-law growth with approximate periodic superpositions do occur within fixed families, neither the power law nor the period are universal (see \cite{ChoHLP03b,Bau18,MarCorRue24}, but see also \cite{GunBauHil24}, as well as \cite{ChoHirLie04,MarCorRue25} for the interesting case of twisting complex scalar fields, for which numerical evidence supports the restoration of universality). Likewise, work in electrovacuum \cite{BauGH19,PerB21} suggests different power laws and periodicities for dipolar and quadrupolar waves, even though higher fine-tuning may lead to the latter being dominated by the former at late times (see \cite{ReiCho23}).  Regardless, there currently is no evidence for a universal and strictly DSS critical solution in this case. Finally, for the collapse of vacuum gravitational waves, as summarized in Table~\ref{Table:PreviousResults}, numerical experiments suggest that, while power-law growth with approximate periodic superpositions do occur within at least some fixed families, neither the power nor the period are universal.

Given this confusing state of affairs one may fear that that certain numerical results are wrong or misleading. This is exacerbated by the very real challenge of providing rigorous convergence testing with adaptive meshes when solutions depend sensitively on the continuum data. Consequently, the
comparison of results with different methods is
important. In~\cite{BauBC23} we compared scaling of the Kretschmann scalar with two different families of centered ($\rho_0=\rho_z=0$) Brill waves (one with a positive amplitude parameter~$A$ in~\eqref{eq:Brill_seed} and the other negative), and found excellent agreement.

To supplement those results we now provide a similar comparison for the time-asymmetric waves based on the ansatz \eqref{eq:Teukolsky_seed}. We follow the bisection steps made in~\cite{LedK21} (with \prague\!) in the upper branch of solutions to
the initial data constraints with the ansatz~\eqref{eq:Teukolsky_seed}
but this time with \bamps. We then performed a handful of additional evolutions slightly closer to the threshold than in the earlier study. A plot of the Kretschmann scalar against the parametric distance from the threshold is given in Fig.~\ref{fig:Scaling}. The agreement between the two methods is very encouraging. The data appear to satisfy a power-law with a periodic wiggle, yet neither the slope nor the period are the same as those from various other families that have been studied. The agreement from the two codes reinforces the impression that the variation of power-laws and periods is not just an artifact of numerical error, and undermines a simple argument for universality.

\begin{figure}[t!]
\includegraphics[width=\linewidth]{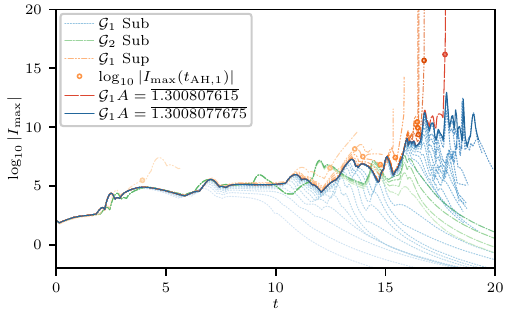}
\caption{\label{fig:Peeling} The maximum of the Kretschmann scalar as a function of coordinate time from the
  \bamps~bisection of time-asymmetric data. We see that the curves stay
  very close to each other until the spacetime `decides' to either
  decay or collapse. There is a small window of amplitudes in which
  the data are subcritical but in which our original
  gauge~$\mathcal{G}_1$ results in coordinate singularities. For these
  spacetimes we worked with a second choice~$\mathcal{G}_2$ (see the
  main text for details). The best-tuned sub- ($A = \overline{1.3008077675}$) and supercritical data ($A = \overline{1.300807615}$)
  are plotted solid blue and long-dashed red lines respectively. In supercritical data the circles denote the
  first times at which we are able to find horizons. In the best-tuned
  data this is essentially as the method begins to fail through the
  explosion of curvature, making further bisection very challenging.}
\end{figure}

In~\bamps~we generally work with a variation of harmonic damped wave gauge employed in binary black hole evolutions (see~\cite{LinSzi09,HilWB16,CorRenRue23} for details). Performing these evolutions made clear one of the shortcomings of working with (our choice of) generalized harmonic coordinates for these applications.  Specifically, our gauge is parametrized by
\begin{subequations}\label{eqn:GHG_gauge}
\begin{align}
  \p_t\alpha
  &=-\alpha^2K + \eta_L\log\big(\frac{\gamma^{p/2}}{\alpha}\big)
  +\beta^i\p_i\alpha\,,\\
  \p_t\beta^i&=\alpha^2\,{}^{(3)}\Gamma^i-\alpha \p^i\alpha
  -\eta_S\beta^i+\beta^j\p_j\beta^i\,,
\end{align}
\end{subequations}
where $p$, $\eta_L$, and $\eta_S$ are free scalar functions of the metric and coordinates. Generally we choose~$p=1$, $\eta_L=0.2$, and $\eta_S=12$, which we will refer to as gauge~$\mathcal{G}_1$ in the following. Far from the threshold, calculations in this gauge proceed straightforwardly, and the same is true close to the threshold, even beyond the best-tuned data from the earlier calculations with~\prague. There is a window of amplitudes of the initial data, however, in which, despite the spacetime being subcritical, the \bamps~calculation fails in this gauge. In particular, after the expected largest peak in the Kretschmann scalar, the data should ultimately disperse.  Instead, shortly after they start to do so, a large, very sharp, probably divergent spike forms in the lapse function close to the position of the peak in the Kretschmann scalar. To overcome this, we experimented with different choices of the above parameters. Ultimately we evolved these spacetimes with~$p=0.5$, which we will refer to as gauge~$\mathcal{G}_2$ in the following, and which dealt well with the dispersion in this troublesome interval. The spacetimes evolved in this gauge are marked with green markers in Fig.~\ref{fig:Scaling}. Frustratingly, however, the adjustment itself fails closer to the threshold, so we cannot simply work with it over the entire range of amplitudes. It may well be that there are such choices, but clearly, from the methods point of view, the first lesson from our experiments is that further investigation is needed.

It is often the case in numerical studies of critical collapse that, when gauge is changed, the effective threshold amplitude will change substantially due to the change in numerical error. Such a perturbation may also affect the maxima of curvature scalars. Since no such perturbation is observed here (the green markers lie exactly at the expected points), we conclude that, in fact, the numerical data are fairly accurate. This is perhaps an artifact of the insufficient tuning, but may also indicate that changing between our gauges~$\mathcal{G}_1$ and~$\mathcal{G}_2$ does not have a large effect on the numerical value of the threshold amplitude (which varies only because of numerical error).

\begin{table}[t]
\centering
  \begin{ruledtabular}
  \begin{tabular}{lc|ccc}
Data & $-\log_{10}(A_\star-A)$ & fit type &$\gamma$  & $\Delta$\\
\hline
      & (all)  & linear & $0.22 \pm 0.01^*$ & -- \\
Brill & $ >  2$  (near) & peaks  &  --              & $0.67 \pm 0.03^{~}$\\
      & $ >  2$  (near) & double & $0.23\pm 0.00^*$ & $0.71 \pm 0.02^*$ \\
\hline
      &  (all)  & linear & $0.37 \pm 0.01^{~}$ & -- \\
      & $ < 4.5$   (far)  & linear & $0.26 \pm 0.01^*$& -- \\
TA    &$ \geq 4.5$ (near) & linear & $0.51 \pm 0.02^{~}$ & --\\
      &$ \geq 4.5$ (near) & peaks  & --              & $^{~}1.1 \pm 0.04^{~}$ \\
      &$ \geq 4.5$ (near) & double & $0.51 \pm 0.01^*$& $1.32 \pm 0.06^*$\\
\end{tabular}
  \end{ruledtabular}
  \caption{Fit parameters~$\gamma$ and~$\Delta$ obtained from Eq.~\eqref{eq:wiggle} for the \bamps~Brill and time-asymmetric data in different regions of parameter space. Our linear fits only estimate~$\gamma$, whereas double fits estimate both~$\gamma$ and~$\Delta$. `Peaks' refers to finding the oscillation periodicities after fitting the linear piece. Values marked with * are plotted in Fig.~\ref{fig:Scaling}.}
\label{Table:Fits}
\end{table}

Coming now to the details of the data in Fig.~\ref{fig:Scaling}, we fit power-laws and, in the relevant regions, a sine wave to both the Brill wave and time-asymmetric families. The values of the exponent~$\gamma$ obtained from a linear fit are presented in column 4, rows 1 and 4 of Table.~\ref{Table:Fits}, where the former is also plotted in Fig.~\ref{fig:Scaling} as~$\gamma_{\mathrm{B,far}}^\mathrm{s}$ with `s' denoting that the fit is to a single variable. Observe that for time-asymmetric data, $\gamma$ is comparable to that found in~\cite{AbeE93} in the scaling of apparent horizon masses for similar data (and, presumably by coincidence, to Choptuik's spherical critical exponent~\cite{Cho93} of~$\sim0.37$ for a massless scalar field). However, very much like examples in~\cite{ReiCho23, MarCorRue25} for the time-asymmetric waves we find that the fit is better split into two regions. These can be found in column 4, rows 5 ($\gamma_{\mathrm{TA,far}}^\mathrm{s}$ in Fig.~\ref{fig:Scaling}) and 6 of Table.~\ref{Table:Fits}. In common with the examples of~\cite{ReiCho23, MarCorRue25} is an approximate factor two between the exponents in the coarser and finer tuned regions. It remains to be seen if a compelling model can be given to explain this behavior.

\begin{table*}[t!]
  \centering
  \begin{tabular}{l|l|l|l|c|c|l|l|c|c}
    \hline \hline
    \multirow{2}{*}{$A$} & \multirow{2}{*}{$M_{\mathrm{ADM}}$}
      & \multicolumn{4}{c|}{\bamps}
      & \multicolumn{4}{c}{\prague}
      \\
    \cline{3-10}
    & & \multicolumn{1}{c|}{$t$} & $M_{\mathrm{AH}}$ &
                    $M_{\mathrm{AH}}/M_{\mathrm{ADM}}$ & nature
    & \multicolumn{1}{c|}{$t$} & $M_{\mathrm{AH}}$ &
                    $M_{\mathrm{AH}}/M_{\mathrm{ADM}}$ & nature \\
    \hline
    \hline
    $\overline{1.25}$ & 0.6752 & ~\,3.9286 & 0.2809 & 0.4160 & \multirow{6}{*}{single}
                      & 11 & 0.367 & 0.5435 & \multirow{7}{*}{single}\\
    $\overline{1.3}$ & 0.5904 & 12.4727 & $0.1804^\dagger$ & 0.3056 &
                     & 15.75 & 0.198 & 0.3353 \\
    $\overline{1.3005}$ & 0.5894 & 13.5739 & 0.2095 & 0.3554 &
                        & 16.5 & 0.172 & 0.2918 \\
    $\overline{1.3006}$ & 0.5893 & 13.9681 & 0.1722 & 0.2922 &
                        & 17 & 0.168 & 0.2851 \\
    $\overline{1.3007}$ & 0.5891 & 14.7541 &  0.1406 & 0.2387 &
                        & 17.75 & 0.13 & 0.2207 \\
    $\overline{1.30076}$ & 0.5890 & 15.4466 & 0.1251 & 0.2124 &
                         & 19 & 0.131 & 0.2224 \\
    \cline{3-6}
    $\overline{1.3008}$ & 0.5889 & 16.4500* & 0.05289 & ~\,0.08981 &
                                              \multirow{6}{*}{bifurcated}
                        & 20 & 0.0818 & 0.1389 \\
    \cline{7-10}
    $\overline{1.3008012}$ & 0.5889 & 16.4680* & 0.04410 & ~\,0.07488 &
                           & 20 & 0.0367 & ~\,0.06232 &
                                                  \multirow{4}{*}{bifurcated}\\
    $\overline{1.3008025}$ & 0.5889 & 16.5000* & 0.03144 & ~\,0.05339 &
                           & 20.25 & 0.0395 & ~\,0.06707 &\\
    $\overline{1.300805}$ & 0.5889 & 16.7582 & 0.01018 & ~\,0.01729 &
                          & 20.25 & 0.0381 & ~\,0.06470 \\
    $\overline{1.3008075}$ & 0.5889 & 20.75 & 0.0285 & ~\,0.04840 &
                           & 17.5460 & 0.01375 & ~\,0.02335 \\
    \cline{7-10}
    $\overline{1.300807615}$ & 0.5889 & 17.7112 & 0.01235 & ~\,0.02097 &
                             & \hfil -- \hfil & \hfil -- \hfil &
                               \hfil -- \hfil & \hfil -- \hfil\\
    \hline \hline
  \end{tabular}
  \caption{Initial horizon masses~$M_{AH}$ from time-asymmetric wave evolutions with \bamps~and \prague. Times marked with * were determined by our earliest AH data output time, rather than the first slice in which we may have been able to find horizons had data been available. Masses and times for \bamps~AHs are quoted from the flow finder as we believe the calculation to be more accurate, except for the spacetime~$A=\overline{1.3}$, marked with~$\dagger$, for which we used the shooting method due to technical issues. For the \prague~data values are quoted from the shooting method. See the main text for further discussion.
  \label{Table:AHs}}
\end{table*}

In the standard picture of critical collapse with a single unstable mode with critical exponent~$\gamma$, if the threshold solution was DSS with period~$\Delta$, its oscillations would leave an imprint on the scaling of the maximum of the Kretschmann scalar~\cite{HodP97,Gun97} as
\begin{align}\label{eq:wiggle}
\log{I_{\text{max}}} &\sim -4\gamma\log(A_{\star}-A)\nonumber\\ 
&\quad + C \sin(-\omega\log(A_{\star}-A) + \delta),
\end{align}
with~`$\sin$' here a stand-in for an arbitrary~$2\-\pi$ periodic function that we use simply to fit to the angular frequency~$\omega=2\pi\gamma/\Delta$. In order to compute the assumed DSS periods from our data, we isolate the regions where the clearest oscillations occur, that is~$\log_{10}(A_{\star}-A) > 2$ for the Brill family and~$\log_{10}(A_{\star}-A) \geq 4.5$ for the time-asymmetric family, then remove the linear contribution. Looking at the peaks of the oscillations we obtain the values presented in column 5, rows 2 and 7 of Table.~\ref{Table:Fits}. The value of~$\Delta$ for the Brill data obtained here is close to the value~$0.60$ provided by~\cite{AbeE94}, but recall that the exponent did not match for this family. As a second approach, we perform a double fit in which we minimize the sum of squared residuals from Eq.~\eqref{eq:wiggle} to fit~$\gamma$ and~$\Delta$ in those regions simultaneously. We use the exponents and periods just described as initial guesses for the double fit, though we observe the result does not depend sensitively on them. The resulting fits are presented in columns 4 and 5, rows 3 and 8 of Table~\ref{Table:Fits} and are also plotted alongside our numerical data in Fig.~\ref{fig:Scaling} with the exponents labeled as~$\gamma_{\mathrm{B,near}}^\mathrm{d}$ and~$\gamma_{\mathrm{TA,near}}^\mathrm{d}$, where the `d' denotes a double fit. Their agreement with the data may suggest structure in the threshold solutions of each family, and could indicate the presence of such approximately DSS threshold solutions.

The different values found for each family in both the period and the exponent are direct evidence of a lack of universality at the vacuum threshold of collapse in axisymmetry up to the current level of tuning. The possibility of these slopes becoming universal closer to the threshold remains open, but we have no evidence for this at present.

To help understand how the dynamics of spacetime is represented in our local time coordinates and how they vary as we vary the amplitude of the initial data changes, we present in Fig.~\ref{fig:Peeling} the maximum of the absolute value of the Kretschmann scalar as a function of time, again from \bamps~data, this time from both sub- and supercritical data. Despite agreement in the values of the maxima, the plots from the codes differ a lot due to the different time coordinates employed. Despite their foliation dependence it is also often expected that, with reasonable time coordinates, as the threshold is approached from the supercritical side, similar scaling will be observed in the AH masses. The extent to which the AHs that we find in these two foliations agree is therefore of interest. As a naive first examination we have marked in Fig.~\ref{fig:Peeling} the times at which we first find AHs in the supercritical data. Generally speaking AHs are found in a neighborhood of the large features in the Kretschmann scalar and, somewhat disappointingly from the point of view of accuracy, we see that this happens after these features begin to explode without bound.

In the centered~$A<0$ Brill wave family from~\cite{BauBC23}, it was found that subsequent peaks in the oscillations of the Kretschmann scalar were occurring closer to the origin. This encouraged the introduction of coordinates adapted to an assumed DSS centered at the origin to study those spacetimes bottom panel of Fig.~2 in~\cite{BauBC23}. From this it was concluded that there appears to be an approximate DSS behavior in the centered~$A<0$ Brill wave family at the threshold. Making the same first steps for the time-asymmetric family, we instead find, within the \bamps~foliation, that the proper distance of these peaks do not accumulate at the origin, similar to what is shown in Fig.~3~\cite{BauBC23}.

As a first step in the comparison of the apparent horizons, we
verified that our two finders give compatible results on identical data. With that done, we made a careful sweep through the data, not only to classify but to find horizons at the earliest instant of time possible. In Table~\ref{Table:AHs} we thus give the values of the
horizon masses~$M_{AH}$ for the supercritical time-asymmetric wave evolutions as calculated from \bamps~and \prague~at the first time slice we were able to find them. In both the generalized harmonic and the quasimaximal foliations the data strongly support a trend towards smaller~$M_{AH}$ that will vanish at the threshold. That said, masses
are not monotonic functions of~$A-A_\star$. The reason
for this is at least partially numerical in nature: either that we have missed the
earliest horizon through output that is too coarse in time, or because
of the challenge of finding them when they are very small. As in
earlier works we find that sufficiently close to the threshold of
collapse the first horizons we find 
form in pairs as a result of reflection symmetry. After the
initial discovery we are generally able to find horizons reliably at
later times, until our data become noisy due to our crude treatment of the post-collapse phase (we do not excise or use a Gamma-driver shift, the natural approaches to handling black holes, with the \bamps~and \prague~codes, respectively). Beyond these qualitative similarities there are substantial quantitative differences between the values that we find for the horizon masses. For instance, we find bifurcated horizons one bisection step earlier in the \bamps~data than the \prague~data, and rather different values (up to a factor of a few) in the horizon masses themselves. This is no contradiction because of the well-known foliation dependence of the apparent horizon, but highlights the difficulty of trying to look for power-law scaling in the initial horizon masses beyond spherical symmetry as we tune. In our best-tuned data we find that the around~$2\%$ of the ADM mass lies inside the black hole around the instant of formation. Because of the differences discussed above, we do not plot the horizon masses obtained with the two codes together as a function of distance from the threshold.

Regarding the two AH finders, we observe that, while the shooting method is more robust and is in principle superior because it does not assume star-shaped horizons, in practice in our current implementation has the disadvantage on \bamps~data of using an intermediate uniform grid rather than the native AMR hierarchy. (And, in fact, within this family of time-asymmetric initial data we never find AHs that are not star-shaped). Because of this down-sampling of the resolution, the shooting method has difficulty finding relatively large horizons close to the moment of formation. In that scenario, the shapes of the AHs are deformed enough that the calculation is delicate. Hence, the resolution needed for accurately covering the larger area becomes tricky to reach. Additionally, if the edge of the AH is close to the edge of \bamps's computational domain, the shooting method also has trouble finding the AH. This can happen for bifurcated AHs close to the bisection point where they first split in the \bamps~domain when using reflection symmetry about the x-axis. These also decrease the advantage of it being a fast and computationally cheap method. For these larger horizons, the flow method fares much better. Even in the cases where we do find relatively large horizons with both methods, the flow method is able to find it in earlier time slices than the shooting method in some cases. On the other hand, as we close in on the critical threshold and have smaller and smaller horizons, the shooting method works much better, as the smaller region is well resolved even at lower resolutions, and it can reliably locate the AHs fast. The flow method struggles here as in our implementation we generally set a maximum value of the expansion as a termination condition for the search, and in practice we find that the expansion becomes very large in a neighborhood of the horizon, where it rapidly decays to vanish as the horizon itself forms. In these cases, the shooting method can sometimes find the AH earlier than the flow method. Interestingly, we are able to consistently find a MITS in essentially all time slices in all spacetimes with the shooting method during the short time interval between trapped surface formation and code failure.

\subsection{Dynamics of specific spacetimes of interest}
\label{sec:Specific}

We now move on to compare data from the best-tuned data from \bamps~and \prague, considering both Brill and time-asymmetric wave families.

\subsubsection{Subcritical spacetimes}
\label{sec:Subcritical}

\begin{figure}[t!]
\includegraphics[width=\linewidth]{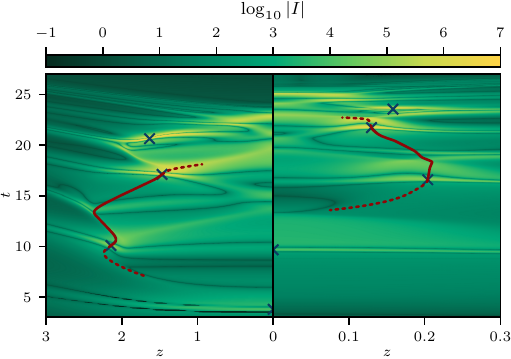}
\caption{\label{fig:geod} Timelike geodesics connecting two neighboring peaks in the Kretschmann scalar for a centered Brill wave with~$A=-3.50909$. In the left panel we plot \bamps~data, and in the right panel those from \prague. The continuous segment of the lines mark the piece of the geodesic that connects two neighboring curvature peaks, and the dashed part their extension beyond. Local maxima in the curvature are indicated by dark blue markers.}
\end{figure}

\begin{figure}[t!]
\includegraphics[width=\linewidth]{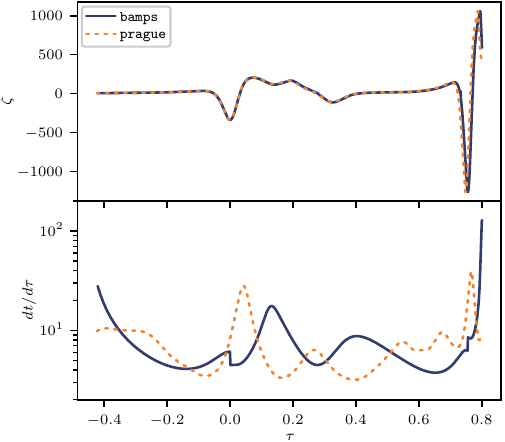}
\caption{\label{fig:zetaatgeod} Top panel: The invariant~$\zeta$ along a geodesic connecting the first two off-center curvature maxima (echoes) for the two simulations of the Brill~$A=-3.50909$ initial data, along the curves marked in Fig.~\ref{fig:geod}. The proper time (on the horizontal axis) is measured with respect to the moment the time-like geodesic passes through the first curvature maximum. Bottom panel: The simulations
  compared in this paper use very different lapse choices. In this
  plot this is illustrated by the rate of change~$dt/d\tau$ of the
  coordinate time along the same geodesic that is considered in the
  top plot. \label{fig:zeta(tau)}}
\end{figure}

The most important discovery of~\cite{LedK21} was the presence of quasi-universal features in twist-free spacetimes close to the threshold. These were unveiled by plotting geometric scalars along geodesics connecting neighboring peaks in the curvature. In
axisymmetry, the scalar
\begin{align} \label{eq:zeta}
    \zeta= \frac{1-\nabla_a\rho\nabla^a\rho}{\rho^2},
\end{align}
with~$\rho$ the circumferential radius, is a geometric invariant. On the symmetry axis, where $\rho = 0$, it satisfies $I = 12\zeta^2$, where $I$ is the Kretschmann scalar (\ref{eq:Kretschmann}). In Fig.~\ref{fig:geod} we plot such geodesics from centered, $A=-3.50909$ Brill wave data according to the native coordinate systems of \bamps~and \prague. As expected, the curves themselves bear only a limited resemblance but, as can be seen in Fig.~\ref{fig:zetaatgeod}, the geometric scalar~$\zeta$ along the geodesic agrees well despite this difference in gauge conditions, so that we recover the profiles observed in Fig.~3 of~\cite{LedK21}.

\begin{figure*}[t!]
\includegraphics[width=\textwidth]{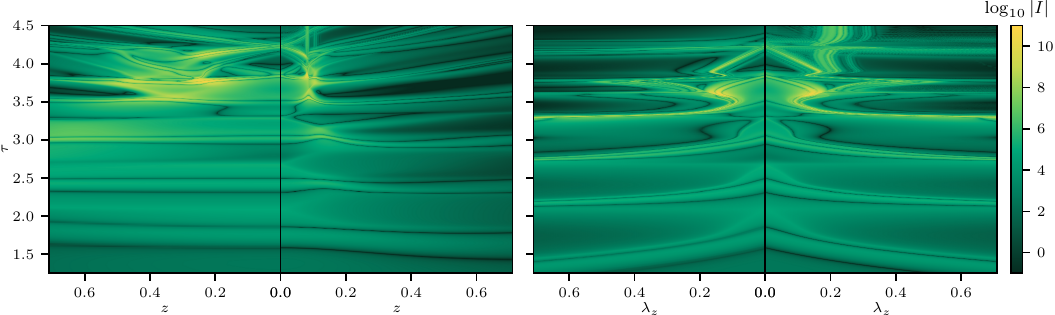}
\caption{\label{fig:TAsinglenull} The Kretschmann scalar~$I$ displayed as a function of proper time~$\tau$ versus~$z$ (left pair) and versus null similarity coordinate~$\lambda_z$ (right pair) for the~$A = \overline{1.30080796}$ time-asymmetric wave, the most tuned common subcritical amplitude for \texttt{bamps} (left in each pair) and \texttt{prague} (right in each pair).}
\end{figure*}

To compare the best-tuned (common) subcritical data from our family of
time-asymmetric waves, we follow the approach employed
in~\cite{BauBC23}. In particular, in the top panel of
Fig.~\ref{fig:TAsinglenull} we plot the Kretschmann scalar in the
local coordinates of \bamps~and \prague~in the symmetry
axis. Naturally, while the range of values is comparable, the data
themselves appear radically different. In the lower panel  of Fig.~\ref{fig:TAsinglenull} we then
introduce future directed single null coordinates, using proper time
at the origin~$\tau$ to label the null-slices and the associated
affine parameter (normalized by~$d\lambda/d\tau=1$ at the origin) to
coordinatize the position on the null-slice itself. The agreement
between the two data sets is remarkable, with differences only
apparent at late times. In contrast to Fig.~2 of~\cite{BauBC23}, here
we did not attempt to introduce coordinates adapted to DSS centered at the origin. The reason for this is that in this family of data, the evidence, discussed above, points against subsequent curvature peaks lying ever closer to the origin. While the threshold solution may be approximately DSS we cannot confirm so with our present post-processing tools.

After the largest peak in the curvature, numerical artifacts are
visible in the \prague~data, which we believe will
converge away with higher resolution (see Fig.~11.6 of~\cite{Khi21}). It is
most likely caused by noise induced at mesh-refinement boundaries,
though the zero-speed characteristic variable in the BSSN constraint
subsystem may also contribute (see~\cite{BerHil09} for discussion of
the latter).  As above, where we saw that it may be valuable to
experiment with different gauges to improve the robustness of the
\bamps~setup, the second lesson is that it may be helpful to try a
different FMR strategy, such as that of
\textsc{CarpetX}~\cite{ShaMoeBra22}, and, perhaps, a formulation of GR
that admits a complete constraint damping scheme such as Z4 or a
conformal decomposition
thereof~\cite{BonLedPal03,GunGarCal05,BerHil09,WeyBerHil11}.

\subsubsection{Supercritical spacetimes}
\label{sec:Supcritical}

We now discuss our best tuned supercritical spacetimes within the
time-asymmetric family~\eqref{eq:Teukolsky_seed}. We begin with a
direct comparison of our {\it common} best-tuned data before giving an
overview of the \bamps~data beyond which our current bisection breaks
down.

\subsubsection*{Comparing best tuned common supercritical spacetimes}

We use double-null coordinates to visualize the agreement between supercritical simulations based on different coordinates. This works only in certain totally geodesic submanifolds because in near-critical spacetimes the null cones in~$3+1$ dimensions develop caustics. Because both the curvature extrema in subcritical evolution and bifurcated horizons of the supercritical spacetimes appear on the~$z$ axis we will consider null geodesics along the~$z$ axis.

For subcritical evolution, one can easily recover the common proper time $\tau_0$ along the central geodesic~$z=0$ and, assuming~$u+v = 2\tau_0$ along the axis, define the double null coordinates by shooting both future and past null geodesics starting at the center. For supercritical spacetimes, the past null cone of the~$z=0$ worldline endpoint within the numerical evolution does not cover important regions of the~$z$ axis submanifold, so that an alternative construction is needed. We chose to start with a future outgoing null geodesic along the~$z$ axis originating from the center at some particular central proper time~$\tau_0^\text{ini}$ and labeled $u=0$. The affine parameter along this geodesic is normalized with respect to the central observer and thus is compatible between simulations. Then, the coordinate~$v$ is defined by this affine parameter assigning these values to the future-pointing {\em ingoing} null geodesics $v=\rm const.$ starting at the cone~$u=0$. When these~$v=\rm const.$ rays reach the origin, these events define values of the coordinate $u>0$ so that we get $u-v=0$ at the center; however, due to coordinate construction, their sum is not necessarily related to~$\tau_0$. With this approach, all (numerically) constructed null geodesics are future-pointing and the $u$ and $v$ coordinates cover the whole interior of the~$u\ge 0$ null cone.

\begin{figure*}[t!]
\includegraphics[width=\textwidth]{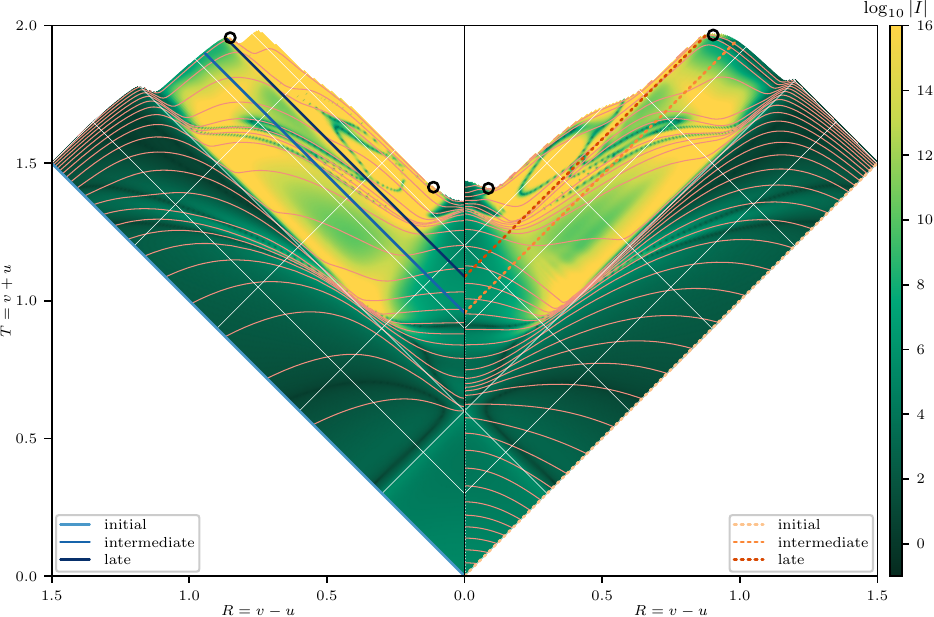}
\caption{\label{fig:KretschDoubleNull} The color-coded Kretschmann scalar~$I$ on the $z$-axis in double null coordinates compared for \bamps~(left) and \prague~(right) codes, $A=\overline{1.3008075}$. The coordinates are started
  from the central worldline at the same proper time~$\tau_0$ and are
  chosen so that, at the center, we have~$u=v$. This construction uses the
  affine parameter along~$u=0$ geodesics as the common coordinate of
  both simulations. Due to the highly dynamical nature of the spacetime curvature, the figure looks significantly different when different values of~$\tau_0$ are chosen (see the main text for definitions). We show profiles of~$I$ along the three null geodesics labeled `initial', `intermediate', and `late' in Fig.~\ref{fig:TAIalongNull} below. The black circles mark the poles of the (bifurcated) AH when it is first detected.}
\end{figure*}

In Fig.~\ref{fig:KretschDoubleNull} we compare the Kretschmann scalar~\eqref{eq:Kretschmann} inside the future null cone~$u=0$ for our common best tuned time-asymmetric data~$A=\overline{1.3008075}$. In this plot, the coordinates~$u$ and~$v$ are both compactified using~$\arctan$, which holds the null cones fixed at~$45$-degrees in the figure. The two data sets appear compatible. Clearly, since the shaded region on the left extends higher in the figure, the \bamps~gauge slicing ventures closer toward the curvature singularity though, unsurprisingly, as curvature scalars diverge towards the top of the `very yellow' region substantial constraint violation is present there. As reported already in Table~\ref{Table:AHs}, in both sets of data we find apparent horizons centered away from the origin, initially with around~$2\%$ and~$5\%$ of~$M_{\textrm{ADM}}$ for the \bamps~and \prague~data, respectively. To give a sense of the degree of foliation dependence of the AHs, we mark their intersection points with the symmetry axis, at the moment when they are first formed, by the black circles in Fig.~\ref{fig:KretschDoubleNull}. These markers appear approximately in the same positions, but, as the data are highly curved in the region they are found, we do not suppose this indicates any contradiction in the masses obtained, particularly since we have verified the AHs in the \bamps~data with both of our two AH finders. Compatibility across the two AH methods is discussed for even more tuned \bamps~data below.

\begin{figure}[t!]
\includegraphics[width=\linewidth]{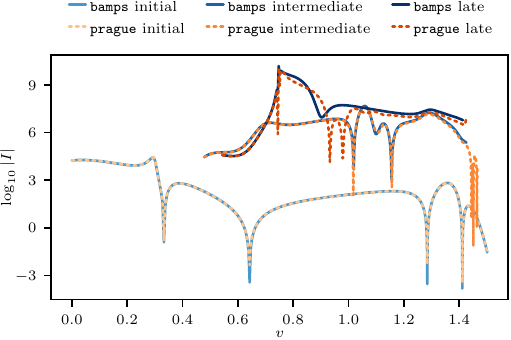}
\caption{\label{fig:TAIalongNull} The Kretschmann scalar~$I$ along the three null geodesics labeled `initial', `intermediate', and `late' in Fig.~\ref{fig:KretschDoubleNull}. At the early and middle time, results from the \bamps~and \prague~codes agree very well, but they show some small differences for the late time.
}%
\end{figure}

In Fig.~\ref{fig:TAIalongNull} we show how the Kretschmann
scalar~$I$ depends on the~$v$ coordinate along several $u=\rm const.$ rays. At early values of the retarded time $u$, the two codes agree very well, but, consistent with our discussion above, they show some small differences at later times.

\subsubsection*{Horizons in our best tuned supercritical spacetime}

\begin{figure}[t!]
\includegraphics[width=\linewidth]{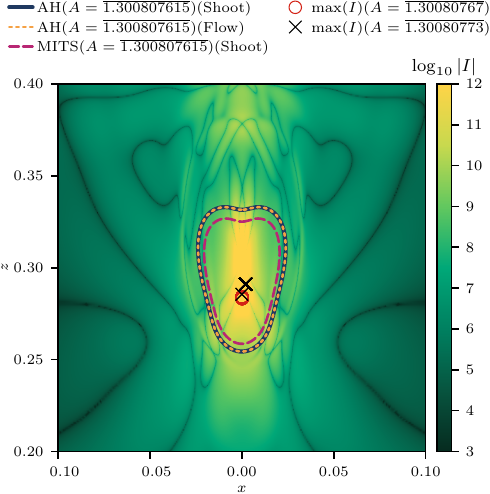}
\caption{\label{fig:TASupTuned} Horizons for the $A =
  \overline{1.300807615}$ time-asymmetric data at $t \sim 17.71115 $, the moment of first AH detection, in the upper half of the~$xz$-plane, as computed with the \bamps~code. There is another symmetric horizon in the lower half, i.e.~for negative~$z$. The colormap shows the log of the Kretschmann scalar~$I$. This is the most tuned supercritical amplitude. Additionally, we show the positions of the maxima of the next two weaker amplitudes that we cannot classify, at a small number of time steps just before they blow up (and are substantially affected by numerical error).}
\end{figure}

As discussed in Sec.~\ref{sec:AHFinders}, we only classify a spacetime as supercritical once we can reliably find AHs. Here we discuss our most fine-tuned classified supercritical spacetime from the time-asymmetric family, corresponding to the amplitude~$A=\overline{1.300807615}$ in the upper branch of the initial data family. This spacetime was simulated only with the \bamps~code. We simulated this spacetime until it crashed due to numerical errors that inevitably occur when the spacetime slices run into (or sufficiently close to) a singularity. We then swept the time slices first with the shooting method AH finder. With this method, the earliest we could find an AH was at coordinate time~$t \sim 17.71115$ (see again Table~\ref{Table:AHs} for an overview). This is shown as  a cross-section of the 2-surface in Fig.~\ref{fig:TASupTuned} with the solid line. We were additionally able to locate a MITS with this shooting method in this same time slice, shown as the dashed 2-surface in the same figure.  We used this AH and took a coordinate 2-spheres that marginally circumscribe it as the initial guess for our flow AH finding method. The resultant AH determined from the flow method found is shown with the dotted line in the same Fig.~\ref{fig:TASupTuned} also. As can be seen, the AHs found by the two methods are essentially identical, thus reinforcing the soundness of the results from the two AH finders. In Fig.~\ref{fig:TASupTuned} we have also plotted the Kretschmann scalar, at the time of first AH detection, as a colormap. As can be seen, the field peaks in exactly the region inside the AH region, as we might have expected.

Confidently classifying the above as our most tuned supercritical spacetime, and~$\ds A = \overline{1.3008077675}$ as the most tuned subcritical, constrains the critical threshold amplitude to within a margin of~$\sim 1.525 \times 10^{-7}$. We fail to classify the spacetimes within this range successfully, as the simulations blow up at a certain point, but we fail to locate horizons. However, in order to investigate these, we have plotted the coordinate positions of the maxima in the Kretschmann scalar for two intermediate amplitudes of~$A =\overline{1.30080767}$ and~$A = \overline{1.30080773}$ in the last few time slices of each respective run. These are also shown in Fig.~\ref{fig:TASupTuned}. As we can see, the Kretschmann scalar blows up exactly where the peak of the last classified supercritical spacetime occurs, which seems consistent. In this context, the dynamics seen across the parameter space in the scaling plot in Fig.~\ref{fig:Scaling} may offer us some insight. As we discussed around \eqref{eqn:GHG_gauge}, our default gauge $\mathcal{G}_1$ for the \bamps~code fails in a small region, and then again starts working past that region. Upon investigating, we found that this occurs as a large high frequency feature in curvature scalars becomes visible, moving through the computational domain in an amplitude interval that~$\mathcal{G}_1$ fails to resolve properly, but our alternative gauge~$\mathcal{G}_2$ manages due to the lower frequencies of these features in the coordinates when using the latter. Since we can see a strong feature building in the domain in this region for the last classified supercritical data, $\ds A = \overline{1.300807615}$, which is also present in the intermediate amplitudes, it is plausible that our specific gauge again fails to resolve it, and a better choice will help us go further. We leave it to a future project to find gauge choices that allow us to tune further towards the critical threshold as the main goal of this paper is to vet the results from different codes for the spacetimes through systematic comparison.

\section{Conclusions}
\label{sec:Conclusions}

The threshold of gravitational collapse in vacuum presents a
geometrically appealing and physically important problem that has
proven far more subtle than first anticipated, with results over the
years painting a sometimes confusing picture. In this work we have
extended the earlier comparison~\cite{BauBC23} of axisymmetric
twist-free gravitational waves to include different families, and have
considered quantities omitted from the discussion there. In
particular, we have compared off-center Brill waves as treated by the
\bamps, \prague~and \sphGR~codes, moving on to evolve the family of
time-asymmetric wave initial data first treated
in~\cite{LedK21}. For reasons of computational cost we employed just
the \bamps~and \prague~codes for the latter. With these, four distinct
families of initial data (centered Brill waves with~$A>0$ and~$A<0$,
one family of off-center Brill wave data, and one family of
time-asymmetric data) have now been tuned with multiple independent
codes. Encouragingly, in all cases the data have proven physically
compatible.

The families of initial data considered here result in isolated curvature peaks centered on the symmetry axis, and we have therefore focused most of our analysis and comparisons to that axis. Evidently, however, features away from the symmetry axis provide additional information about the global properties of any threshold solution (see for example~\cite{BauGunHil26}).

Numerical work in critical collapse lives and dies by the extent to
which we can tune to the threshold of collapse. While we have made
good progress in this, it is still the case that non-vacuum studies,
with or without spherical symmetry, tend to achieve better fine-tuning. An
important purpose of the present work was therefore to identify
avenues for improvement with each of our codes. The first, very clear,
lesson we have learned is that for \bamps~it will be valuable to
systematically investigate different gauge conditions. We very
explicitly found a window in solution space in which coordinate
singularities formed with one of our go-to choices. The second lesson
applies to \prague, where we found that at late times numerical
artifacts are visible in the data that are absent in the equivalent
\bamps~setup. We expect that this is caused by refinement boundaries
and, beyond this, could perhaps be mollified by experimenting with
different formulations of GR. (Though not apparent from the comparison
here, for \sphGR~a possible algorithmic improvement for data with
multiple centers of collapse would be to employ cylindrical polar
coordinates~\cite{ReiCho23}).

The comparison furthermore made clear the value of having multiple
tools to achieve the same task. There were, for instance, cases in
which we would have failed to find apparent horizons using the flow
method without having a first guess provided by our shooting
method. But it was still helpful to use the flow method because it
gives more accurate results on \bamps~data. A third specific lesson is
that we could do a better job at postprocessing strong-field and highly dynamical vacuum data. For example, when finding AHs neither of our two methods is efficient, or robust enough to be used without hand-holding from the user. A second example is that in the data considered in~\cite{BauBC23} it was fortunate that in one of the families we considered (centered~$A<0$ Brill waves) it appeared that subsequent curvature peaks were accumulating closer and closer to the origin, which allowed us to meaningfully introduce coordinates adapted to an assumed DSS centered at the origin. Because of this we discovered that the data were approximately self-similar. In the time-asymmetric data treated here that is not the case and so, even if the data do exhibit such a symmetry, we lack a simple geodesic from which to bootstrap the coordinate construction. A near-term goal is therefore to generalize our earlier setup to allow for the construction of such coordinates more systematically using an arbitrary geodesic. Returning to the subject of AHs, we emphatically learned that, due to the foliation dependence of apparent horizons, scaling properties in the vicinity of the threshold can be assessed more reliably for subcritical data than for supercritical data.  One option would be to evolve supercritical data longer, until the spacetime has settled down into an equilibrium solution and the AH coincides with the event horizon.  However, in addition to significant computer time, this would require gauge conditions better suited for the evolution of black holes than the ones that we employed in this study.

Moving away from methodological questions to physics, our study has reinforced the conclusion of~\cite{BauBC23} and, more broadly, the general consensus of the body of work summarized in the bottom half of Table~\ref{Table:PreviousResults}. In brief; to the extent that we can presently tune, numerical evidence suggests the existence of distinct solutions at the threshold of black hole formation. These may be associated nearby in solution space, via tuning, with (non-universal) oscillatory power-laws in curvature scalars. At the threshold itself, solutions may or may not exhibit approximate self-similarity. In due course we hope to be able to provide data
tuned even closer to the threshold to put these statements on a more solid footing.

\acknowledgments

We are grateful to Carsten Gundlach, Sarah Renkhoff and Isabel Su\'arez Fern\'andez for helpful discussions.

TWB and DH gratefully acknowledge support through the Oberwolfach
Research Fellows program at the Mathematisches Forschungsinstitut
Oberwolfach in 2022 and 2025. DC\@ acknowledges support from the Leverhulme Trust Early Career Fellowship ECF-2025-424. This work was partially supported by FCT
(Portugal) through grant Numbers UID/00099/2025, UID/PRR/00099/2025
and 2023.12549.PEX.

Numerical simulations were performed at the
Leibniz Supercomputing Centre (LRZ) [supported by projects~pn34vo
  and~pn36je], the Deucalion supercomputer at the Minho Advanced
Computing Center (MACC) [with FCT/FCCN projects~2025.09520.CPCA.A3
  and~2025.09510.CPCA.A1], the Bowdoin Computational Grid,
`e-Infrastruktura CZ' computer grid (project No. e-INFRA LM2018140), by the Charles University Project GA UK No. 1176217, Czech Science Foundation Project No.  GACR 21-11268S, and by National Science Foundation (NSF) grant PHY-2308821 to Bowdoin College.

\bibliography{references}

\end{document}